\documentclass[final,5p,times,twocolumn]{elsarticle}
\usepackage{amssymb}
\usepackage{amsmath}
\usepackage{booktabs}

\journal{Theoretical Biology and Medical Modelling}

\begin{document}

\begin{frontmatter}
\title{Boolean network-based model of the Bcl-2 family mediated MOMP regulation}
\author{Tomas Tokar, Zdenko Turcan, Jozef Ulicny}
\address{Department of Biophysics, University of P. J. Safarik, Jesenna 5, 040 01, Kosice, Slovakia}

\begin{abstract}
Mitochondrial outer membrane permeabilization (MOMP) is one of the most important points,
in majority of apoptotic signaling cascades.
Decision mechanism controlling whether the MOMP occurs or not, is formed by an interplay between 
members of the Bcl-2 family.
To understand the role of individual members of this family within the
MOMP regulation, we constructed a boolean network-based mathematical model of 
interactions between the Bcl-2 proteins.
Results of computational simulations reveal the
existence of the potentially malign configurations of activities of the Bcl-2 proteins,
blocking the occurrence of MOMP, independently of the incoming stimuli.
Our results suggest role of the antiapoptotic protein Mcl-1 in relation 
to these configurations.
We demonstrate here, the importance of the Bid and Bim according 
to activation of effectors Bax and Bak, and the irreversibility of this activation.
The model further shows the distinct requirements for effectors activation, 
where the antiapoptic protein Bcl-w is seemingly a key factor preventing the 
Bax activation.
We believe that this work may help to describe
the functioning of the Bcl-2 regulation of MOMP better, and hopefully
provide some contribution regarding the anti-cancer
drug development research.

\end{abstract}


\end{frontmatter}

\section{Introduction}
\label{introduction}

Apoptosis is a process of programmed cellular death, distinct from necrosis
~\cite{ulukava_2011,wyllie_2010}, 
which can be well distinguished by its morphology~\cite{elmore_2007}.
It is an important homeostatic mechanism, defects of which may cause
variety of serious diseases, including the neurodegenerative disorders
~\cite{mattson_2006}, autoimmune diseases~\cite{nagata_autoimmune_2010}, 
or even a cancer~\cite{burz_apoptosis_2009,fulda_2009,strasser_deciphering_2011}.
Signals leading to an apoptosis initiation can originate from an extracellular 
environment or from a cell's internal space~\cite{strasser_apoptosis_2000,
strasser_deciphering_2011}.
Apoptotic signals further proceed through an apoptotic signaling and regulatory 
network, that contains several control points
~\cite{strasser_apoptosis_2000,strasser_deciphering_2011}.
One, highly important of such points is formed by 
a family of Bcl-2 (B-cell lymphoma 2) proteins~\cite{danial_2004, chipuk_2008}. 
An interplay between the Bcl-2 family's members controls one of 
the most crucial apoptotic events - the mitochondrial outer membrane 
permeabilization (MOMP)~\cite{tait_2010,landes_2011}. 

MOMP allows release of apoptotic key players - Smac/DIABLO and a cytochrome c, 
from a mitochondrial intermebrane space to a cytosol~\cite{tait_2010,landes_2011}.
In presence of ATP, released cytochrome c binds to a cytosolic protein Apaf-1, 
causing Apaf-1 oligomerization and a recruitment of an inactive pro-caspase-9, 
leading to formation of a multi-protein complex known as an apoptosome 
~\cite{mace_2010,perez-paya_2011,kulikov_2012}.
Within the apoptosome, pro-caspase-9 subsequently undergoes processing and 
activation~\cite{mace_2010,perez-paya_2011,kulikov_2012}.
The active caspase-9 proteolytically activates caspase-3~\cite{wurstle_2012}.
Smac/DIABLO, once released to the cytosol, inhibits XIAP 
(X-linked inhibitor of apoptosis) - the most prominent suppressor
of caspases -3 and -9~\cite{martinez-ruiz_2008}.
Caspase-3 and other effector caspases (caspases -6 and -7) 
are the primary executioners of the apoptosis~\cite{olsson_2011, strasser_deciphering_2011}. 
Activation of these makes the point of no-return, 
after which apoptosis irreversibly occurs~\cite{green_1998}.
Although, mitochondria-independent apoptotic signaling pathways 
are currently well known~\cite{wallach_2008}, 
the mitochondrial (also known as intrinsic) 
pathway currently remains to be considered as the major one~\cite{estaquier_2012}.

The Bcl-2 family mediated control of MOMP is carried in "all or nothing" manner, 
giving no possibility of intermediate MOMP.
A mechanism of controlling such a decision resembles what a complex systems 
science describes as a bistable switch.
Currently the steady-state bistability is considered as the most suitable 
framework for a description of the apoptosis control. 
This interesting system properties have made Bcl-2 family an 
attractive subject of a mathematical modeling and computer simulations.
There are several works regarding a modeling and a simulation of 
the Bcl-2 family and the control of MOMP, 
revealing and examining a variety of non-linear system behaviors such as 
robustness, stimulus-response ultrasensitivity
~\cite{chen_robustness_2007} 
and bistability~\cite{cui_two_2008,sun_2010,tokar_2012}. 
Besides these, Bcl-2 family was involved in several other, 
more general models of apoptosis signaling
~\cite{bagci_2006,albeck_quantitative_2008,harrington_2008}.

All the above-mentioned models were utilized to dynamically 
simulate the chemical reaction kinetics of the studied biological system. 
Complexity of such models is often reduced by grouping 
of several functionally similar species together. 
Moreover, the most prominent group's member is taken as 
the model's representation of the whole group of species. 
Although previous models of Bcl-2 regulatory network are on 
the various levels of details, they all adopt such simplification.
This is done usually by grouping the Bcl-2 family's members into three groups
according to their structural and functional classification. 
Such division provides reasonable trade-off between 
model's simplicity and plausibility. 
However, when grouped together, certain important 
functional specificities of Bcl-2 family's individuals are ignored.

The critical limitation in the development of the detailed, 
quantitative model of the Bcl-2 family is the absence of biological data. 
The identification of the quantitative parameters 
(e. g. reaction rates and/or physiological 
concentrations of species) is dependent on the relevant experimental
\textit{in vivo} measurements, which are still a systems biology bottle neck
~\cite{schlatter_2009}. In absence of quantitative data, such detailed models of
Bcl-2 family must be limited to a qualitative manner. 

In this work we provide a qualitative model of the Bcl-2
family mediated regulation of MOMP based on the Boolean network modeling.
The boolean network (BN) approach is one of the most suited to
a qualitative modeling of complex biological systems, 
even with a limited biological knowledge
~\cite{schlatter_2009,mai_boolean_2009}.
BN, first introduced in the late 1960s~\cite{kauffman_1969}, 
has been originally used to model gene regulatory networks and 
signaling pathways~\cite{helikar_2011}. Although, BN 
does not model dynamics of the studied system, it may reveal and examine
many interesting complex-system properties~\cite{helikar_2011}.
The certain members of the Bcl-2 family appeared for the first time in BN-based model
in work of Calzolari et al~\cite{calzolari_2007}, 
involved in the model of an apoptosis gene network. 
Mai and Liu~\cite{mai_boolean_2009} and few months later Schlatter et al
~\cite{schlatter_2009}, published currently the most recent BN-based 
models of apoptosis, 
both containing simplified mechanism of Bcl-2 family MOMP control.
However, as far as we know, there haven't been published any BN-based model, 
focused solely on the Bcl-2 family.

\section{Materials and methods}
\label{materials_methods}

\subsection{Model's structure and its biological relevance}
\label{model's_structure}

Bcl-2 family's members are functionally classified as either antiapoptotic, 
or proapoptotic. Structurally, Bcl-2 proteins can be categorized according to 
the number of Bcl-2 homology domains (BH) in their $\alpha$-helical regions
~\cite{elkholi_2011,strasser_deciphering_2011}.
Antiapoptotic members (Mcl-1, A1, Bcl-xL, Bcl-2, Bcl-w and Bcl-B) 
are characterized by the presence of four BH domains (BH1-4)
~\cite{chipuk_reunion_2010,placzek_2010}.
Their role is to prevent MOMP by inhibition of proapoptotic family members
~\cite{chipuk_reunion_2010,placzek_2010}.
Proapoptotic members can be
divided to BH3-only proteins and multidomain proteins - effectors
~\cite{strasser_deciphering_2011}. 
BH3-only proteins can be further subdivided based upon their 
role in apoptotic signaling. 
BH3-only subgroup members, termed sensitizers 
(Noxa, Bad, Puma, Hrk, Bmf and Bik), can only bind to 
antiapoptotic Bcl-2 proteins, forming inactive dimers
~\cite{elkholi_2011}.
Members of another BH3 subgroup, termed activators (Bim and Bid), can act in the same 
way~\cite{elkholi_2011}, 
but in addition, activators can directly activate effectors
~\cite{chipuk_reunion_2010,westphal_2011}.
Effectors, once activated, undergo oligomerization and form pores in 
mitochondrial outer membrane (MOM), leading eventually to MOMP.
~\cite{dejean_2010,landes_2011}.
Therefore, effectors are primary target of inhibition by their antiapoptotic relatives
~\cite{westphal_2011}. 

Altogether, interactions between Bcl-2 family members can be classified into
only three types: 
i) Binding and mutual inhibition between antiapoptotic and BH3-only
proteins. ii) Binding and mutual inhibition between antiapoptotic proteins
and effectors. iii) Activation of effectors by BH3 only proteins.
However, the situation ceases be so simple when we focus on the 
interaction between individual molecules. E.g., the BH3 only sensitizer Noxa
can bind to and inhibit only two antiapoptotic proteins 
(see Table~\ref{tab:bcl2_family_list})
~\cite{chen_differential_2005,elkholi_2011,strasser_deciphering_2011}, but
the other BH3 only sensitizer, 
Puma is able to inhibit five of six major antiapoptotic proteins
~\cite{chen_differential_2005,elkholi_2011,strasser_deciphering_2011}.
On the other hand, while it seems that antiapoptotic protein Bcl-B is
not bound and inhibited by any of the BH3 only proteins
~\cite{rautureau_2010}, the other anti-apoptotic protein 
Bcl-xL is bound by seven of them
~\cite{chen_differential_2005,strasser_deciphering_2011}.
There is also a strong asymmetry in the level of inhibition of effectors by 
antiapoptotic proteins. While Bak is inhibited only by three 
antiapoptotic proteins, Bax is inhibited by all six of them
~\cite{elkholi_2011,strasser_deciphering_2011}.

Knowledge about interaction between Bcl-2 family's members, 
was translated to the boolean-based model we present here. 
The model contains 14 nodes, representing the Bcl-2 family's members.
Each member of the Bcl-2 family is represented by one of the 
model's nodes. Except of the proteins Bad and Bmf that are 
coupled together and represented by a single node, 
and similarly the proteins Hrk and Bik.
The proteins that were coupled together, 
share the same intra-familiar interaction
profiles (see Table~\ref{tab:bcl2_family_list}).
The model contains 34 connection between nodes, 
each representing one molecular
interaction.

\begin{center}
\begin{table*}[ht]
	  \footnotesize
		  \centering
		  \hfill{}
		  \begin{tabular}{lllll}
			  \toprule
			  \textbf{Bcl-2} & \textbf{Bcl-2} & \textbf{Full name of the protein} & \textbf{Binds to and inhibits} & \textbf{Ref}.\\
			  \textbf{family class} & \textbf{family member} & &  & \\
			  \hline
			  \textit{Antiapoptotic} & & & & \\
			  \textit{members:} & & & & \\
			  & Mcl-1 & Myeloid cell leukemia sequence-1 & Noxa, Bim, Puma, Bax, Bak & ~\cite{chen_differential_2005,strasser_deciphering_2011} \\
			  & Bcl-2 & B-cell lymphoma 2 & Bad, Bim, Puma, Bmf, Bax & ~\cite{chen_differential_2005,strasser_deciphering_2011} \\
			  & A1 & Bcl-2 related protein & Noxa, Bim, Puma, Bid, Hrk, Bik, Bax, Bak & ~\cite{chen_differential_2005,strasser_deciphering_2011} \\
			  & Bcl-xL & Bcl-2-like & Bad, Bim, Puma, Bid, Hrk, Bmf, Bik, Bak, Bax & ~\cite{chen_differential_2005,strasser_deciphering_2011} \\
			  & Bcl-w & Bcl-2-like-2 & Bad, Bim, Puma, Bid, Hrk, Bmf, Bik, Bax & ~\cite{chen_differential_2005,strasser_deciphering_2011} \\
			  & Bcl-B & Bcl-2-like-10 & Bax & ~\cite{rautureau_2010}\\
			  \textit{BH3-only} & & & & \\
			  \textit{members:} & & & & \\
			  & Noxa & Phorbol-12-myristate-13-acetate-induced protein 1 & Mcl-1, A1 & ~\cite{chen_differential_2005,elkholi_2011,strasser_deciphering_2011} \\
			  & Bad & Bcl-2 antagonist of cell death & Bcl-xL, Bcl-w, Bcl-2 & ~\cite{chen_differential_2005,elkholi_2011,strasser_deciphering_2011} \\
			  & Bim & Bcl-2like-11 & Bcl-xL, Bcl-w, Bcl-2, Mcl-1, A1 & ~\cite{chen_differential_2005,elkholi_2011,strasser_deciphering_2011} \\
			  & Puma & Bcl-2-binding component-3 & Bcl-xL, Bcl-w, Bcl-2, Mcl-1, A1  & ~\cite{chen_differential_2005,elkholi_2011,strasser_deciphering_2011} \\
			  & tBid & truncated BH3-interacting domain death agonist & Bcl-xL, Bcl-w, A1 & ~\cite{chen_differential_2005,elkholi_2011} \\
			  & Hrk & Harakiri & Bcl-xL, Bcl-w, A1 & ~\cite{chen_differential_2005} \\
			  & Bmf & Bcl-2-modifying factor & Bcl-xL, Bcl-w, Bcl-2 & ~\cite{chen_differential_2005,elkholi_2011} \\
			  & Bik & Bcl-2-interacting killer & Bcl-xL, Bcl-w, A1 & ~\cite{chen_differential_2005} \\
			  \textit{Effectors:} & & & & \\
			  & Bak & Bcl-2-antagonist/killer-1 & Bcl-xL, Mcl-1, A1 & ~\cite{elkholi_2011,strasser_deciphering_2011} \\
			  & Bax & Bcl-2-associated X protein& Bcl-xL, Bcl-w, Bcl-2, Bcl-B, Mcl-1, A1 & ~\cite{elkholi_2011,strasser_deciphering_2011} \\
			  \toprule
		  \end{tabular}
		  \hfill{}
		  \caption{Binding and inhibition between individual members of the Bcl-2 family.}
		  \label{tab:bcl2_family_list}
\end{table*}
\end{center}

\subsection{Transition rules}
\label{transition_rules}

\begin{figure}
	\begin{center}
		\includegraphics[width = 0.4\textwidth]{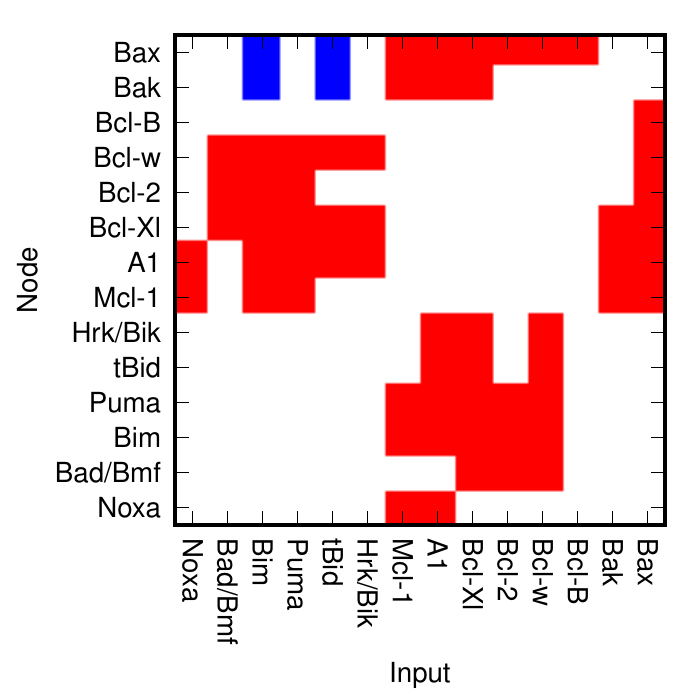}
	\end{center}
	\caption{Relationships between nodes of the model. 
		 The red squares represent the negative relationships of 
		 the inputs toward the related nodes. 
		 Negative relationship corresponds to mutual 
		 binding and inhibition bewteen two members of 
		 the Bcl-2 family (see Table~\ref{tab:bcl2_family_list}).
		 The blue squares represent the positive relationships of
		 the inputs toward the related nodes. 
		 Positive relation corresponds to activation of 
		 the effectors Bax/Bak by certain BH3 only proteins - activators 
		 (tBid, Bim)~\cite{chipuk_reunion_2010,placzek_2010}.}
	\label{fig:relationships}
\end{figure}

In a given time, each of the model's nodes can be in either the active
or inactive state. Each of the nodes is affected by received inputs
from one or several other upstream nodes. 
The state of the node $i$ in the next 
time step $s_i(t + 1)$ is defined by the following transition rule:
	\begin{eqnarray} \label{eq:transition_rule}
		\centering
		s_i (t + 1) &=& \Delta\,\big(e_i + \sum_j r_{ij}\,s_{j}(t) \,\big), \\
		\Delta (x) &=&
		\left\{ 
		\begin{array}{rc} 
			1  \,\,,  &   \,\,  x > 0, \\
			s_i(t) \,\,, & \,\, x = 0, \\
			0  \,\,,  &   \,\,  x < 0
		\end{array} 
		\right.\,
	\end{eqnarray}
Here, $r_{ij}$ specifies the relation of the $j$-th node to $i$-th node, and 
may have three possible values: $r_{ij} = 1$ if $j$-th node activates $i$-th node, $r_{ij} = -1$ if
$j$-th node inhibits $i$-th node and $r_{ij} = 0$, if nodes $j$ and $i$ are not connected
(the relationships between each of the model's nodes
are depicted in the Figure~\ref{fig:relationships}).
The value of $e_i$ defines
the expression of the protein represented by the $i$-th node 
(see Section~\ref{sec:expressions}).
Since Bcl-2 family members inhibit each other by mutual 
binding and formation of inactive 
dimers, our model treats the inhibitory 
relationships between two nodes as bipartite 
(if $r_{ij} = -1$, then $r_{ji} = -1$). 

During the model's simulation, the states of all nodes 
are step-wise simultaneously reevaluated according to the transition rule 
described by the eq. (\ref{eq:transition_rule}), until the simulation is 
terminated (see~\ref{sec:ending_states}).

\subsection{Influence of the external conditions}
\label{sec:expressions}

The transition function of the $i$-th node is dependent on the value of $e_i$.
The value of vector $E$ ($E = \{e_1, e_2,\dots e_{16}\}$) 
represents here, what we termed the "expression" of the the Bcl-2 family proteins.
The value $e_i = 1$, corresponds to the cellular conditions 
allowing the synthesis and, if required, 
the post-translational/post-transcriptional activation 
(e. g. activation of Bid requires proteolytic cleavage by Caspase-8
~\cite{kantari_2011}) of the $i$-th protein.
Alternatively, the value $e_i = 0$, corresponds to the conditions preventing
the synthesis and/or post-translations activation of the $i$-th protein.

Since the model contains 14 nodes, the vector of expressions 
could have $2^{14} = 16384$ of possible values.
However, since other than BH3-only mediated activation of Bax/Bak 
is irrelevant in our work (the subject of our study is
the Bcl-2 family mediated regulation of MOMP), we exclude here 
the expression vectors where $e_{Bax} = 1$ or $e_{Bak} = 1$, 
reducing thus the number of possible values to $2^{12} = 4096$. 
The value of the vector $E$ remains constant during each simulation span.

\subsection{Model's ending states}
\label{sec:ending_states}

The simulation is terminated in the time $t$ if state of 
the model $S$ ($S = \{s_1, s_2,\dots s_{14}\}$) is satisfying
the following condition:
\begin{equation} \label{eq:termination_cond}
        S(t) = S(t - n)\, ,\qquad n = 1, 2, \,\dots 14
\end{equation}
The condition described in the eq. (\ref{eq:termination_cond}),
can imply either that the model converged to 
the steady-state ($n = 1$), 
the model is oscillating between two different states ($n = 2$), 
or the model is periodically orbiting through the set of states ($n > 2$).

The state $S(t)$, satisfying the condition (\ref{eq:termination_cond})
is the model's ending state $S(t_{end})$.
If the $S(t_{end})$ involves the states of both effectors,
$s(t_{end})_{Bax} = 0$ and $s(t_{end})_{Bak} = 0$, then the
$S(t_{end})$ is denoted as the "survival" state.
If the $S(t_{end})$ involves the states of one of the effectors,
$s(t_{end})_{Bax} = 1$, or $s(t_end)_{Bak} = 1$, then the
$S(t_{end})$ is simply denoted as the "pro-momp" state.

\section{Results}
\label{results}

\subsection{We have revealed
		    1046 of the "survival" states, in which the model 
		    can persist while the effectors Bak and Bax
		    remain inactive.}

The very first step, in order to investigate the properties of the model,
was to find the ending states in which model is allowed to persist without
the activation of effectors (Bak and Bax) - survival states 
(see~\ref{sec:ending_states}). 

Therefore, for each of the $4096$ expression vectors 
(see~\ref{sec:expressions}) we performed, 
$4096$ simulations spans, each span starting from one of the
$16384$ of the initial states ($16384 = 2^{14}$, that is the 
number of possible initial states).

We have identified $1046$ of 
the unique survival states. 
The $388$ of these states are logical steady-states, 
remaining $678$ of the survival states are oscillating.
Hereafter, we assume that 
under normal cellular conditions the 
Bcl-2 regulatory mechanism can persist only within
one of these 1046 states.
In any of these states, the model may persist only under
certain expression vectors. 
As we change the expressions, the model may relocate from 
the given survival state to the opposite one.

In the next step, we have investigated the transitions from the survival 
states to other ending states.
To analyze this transitions, for each of the $1046$ of the survival states, 
we have performed $4096$ simulation spans, each span using one
of the expression vectors. 
For each simulation, the given survival state has been used as the initial state of the model.

Around $70\%$ of the $4.2$ millions ($1046\times4096$) of 
simulations led to survival, 
remaining $30\%$ of the simulations led to pro-momp ending states,
where at least one of the effectors was found active.

\subsection{We discovered 200 of the "tumor" states --
		    once accessed by the model, the model remains trapped,
		    unable to turn out, to any functionally distinct state.}

During the analysis of the transitions between the survival states and 
the pro-momp states, we have revealed interesting finding.  
We have discovered, the existence of $200$ survival states, 
from which the model is unable to be turned to pro-momp states, 
regardless of the vector of expressions. 
Moreover, the model, once located in such a "trapping" state, 
can only be turned to other trapping state.
The trapping of the Bcl-2 regulatory mechanism in one of these states, 
would cause fatal malfunctioning of the momp regulation, 
that could possibly result in a tumorgenesis.
Therefore, we denoted these states as the "tumor" states.

These states (see the black squares in the Fig~\ref{fig:states}, top)
can be characterized by strong imbalance in activity of
antiapoptotic vs BH3-only proteins. The tumor states are
especially abundant among the survival states
involving the activity of Mcl-1. 
Although the activity of Mcl-1 is not 
necessary, nor sufficient condition to classify a given model's state as
tumor state, it seems that Mcl-1 plays a certain role here.

\subsection{There are two functionally distinct subsets of 
		    survival states. Those which allow model to activate
		    the Bak, but not Bax and the states allowing activation of
		    both effectors.}

\begin{figure}
 	\begin{center}
		\includegraphics[width = 0.5\textwidth]{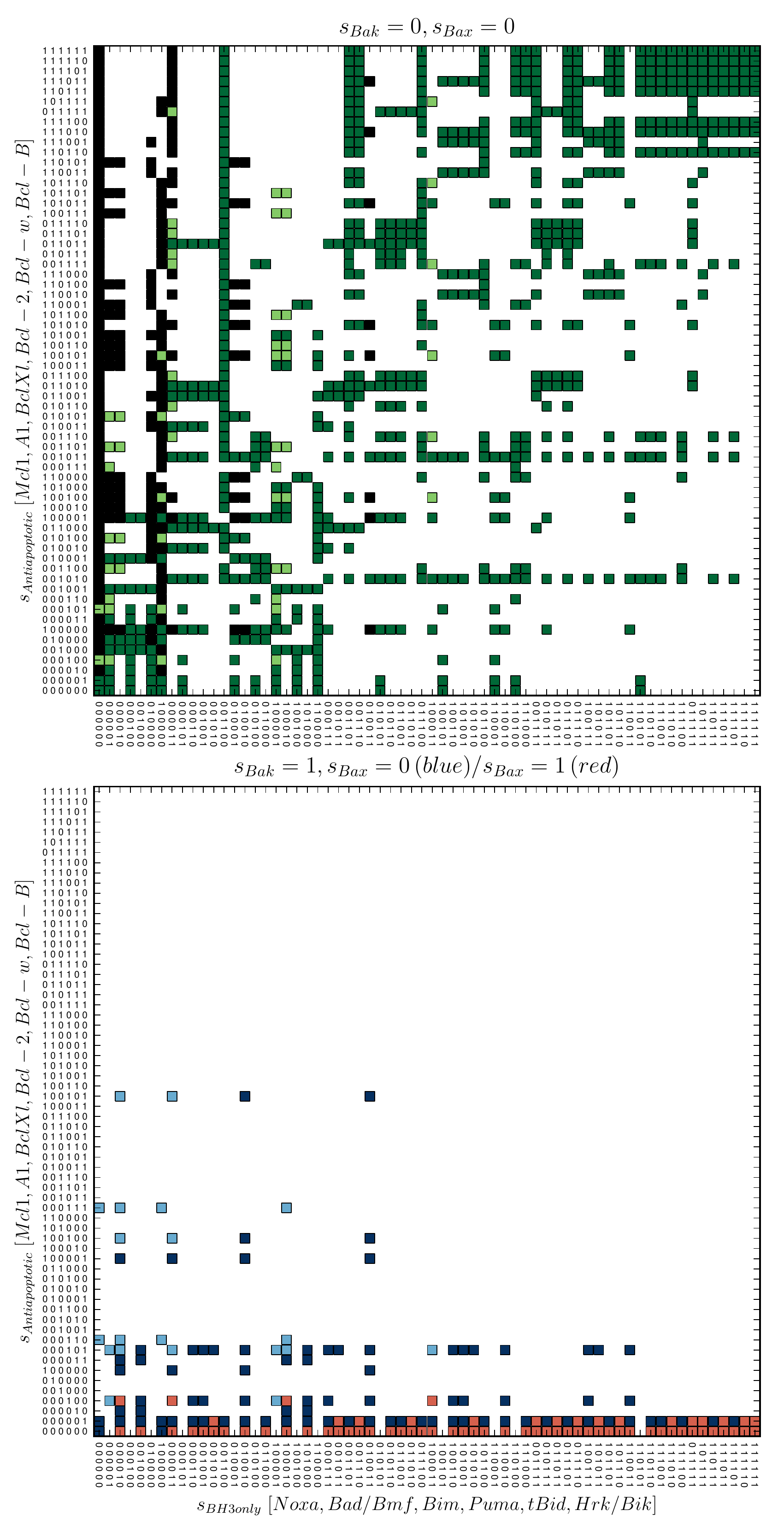}
	\end{center}
	\caption{Model's ending states. The squares are depicting the ending states of the model within the ''configuration space", where
		 the configurations of states of nodes representing the Antiapoptotic and BH3-only proteins are arranged along y- and x-axis, respectively. 
		 The figure on top, depicts the survival states -- Bax, Bax representing nodes are inactive. 
		 The black squares represent the "tumor" states, while the green squares represent the functionally 
		 ''semioptimal" (light green) and "optimal" (dark green) survival states.
		 While the first mentioned, allow only the activation of Bak (transition $T_4$ in Fig~\ref{fig:transitions}), but not Bax,
		 "optimal" states allow activation of both effectors.
		 The figure at the bottom depicts the non-survival states -- either the Bax, or Bak is active (blue squares), 
		 or both of them are active (red squares).
		 The dark blue squares represent those states, which allow subsequent activation of Bax, initiated by the change in the expressions. 
		 In contrast, the states represented by the light blue squares suffer from the inability to allow the activation of Bax.}
	\label{fig:states}
\end{figure}

Remaining $846$ survival states, can further be classified in two groups. 
From the $54$ survival states of the first group, model can relocate
only to pro-momp states where the only active effector is Bak.
From the $792$ survival states of the second group, model can be turned to
states with a single effector (Bak) activity, as well 
as to the pro-momp states, where both effectors are active.
While, the first group we denoted as "semioptimal"
(the light green squares in Fig~\ref{fig:states}, top), 
the second one we denoted as "optimal" 
(the dark green squares in Fig~\ref{fig:states}, top).

Similarly, we may distinguish several functionally distinct subsets among 
the pro-momp states. 
Firstly, the every pro-momp state may be classified according to the activity of
effectors. 
We have found $108$ of the pro-momp states, in which, the Bak is being active, 
but Bax remains inactive
(the blue squares in Fig~\ref{fig:states}, bottom).
Besides these, we also have found $132$ of the pro-momp states, in which both
effectors are being active (the red squares in Fig~\ref{fig:states}, bottom). 
However, we haven't found any such ending state, where the Bax was active, 
while the Bak not, indicating that such state is unaccessible by the model,
regardless of expressions or initial conditions.

Secondly, the first of the mentioned groups -- Bak-active only, 
can further be divided in two functionally distinct subgroups:
states which allow additional activation of Bax 
(the light blue squares in Fig~\ref{fig:states}, bottom), 
and those which doesn't (the dark blue squares in Fig~\ref{fig:states}, bottom).

It is very interesting that, while the first subgroup is accessible only from the 
"optimal" survival states, the second one, can be accessed from both,
"optimal" and "semioptimal" survival states.

\subsection{Survival to momp transition is irrevesible.}

We have performed another series of simulations, 
repetitively simulating the model under each of the 
expressions vectors, while using the pro-momp states as the 
initial conditions. 
We have found out that it is impossible to turn the model from any of the
pro-momp states back to the survival one, regardless of the 
expressions. 

\subsection{The transitions from the survival to momp
		    are caused by expression changes
		    of four distinct types.}
\label{sec:transitions_survival_momp}

\begin{figure}
	\begin{center}
		\includegraphics[width = 0.5\textwidth]{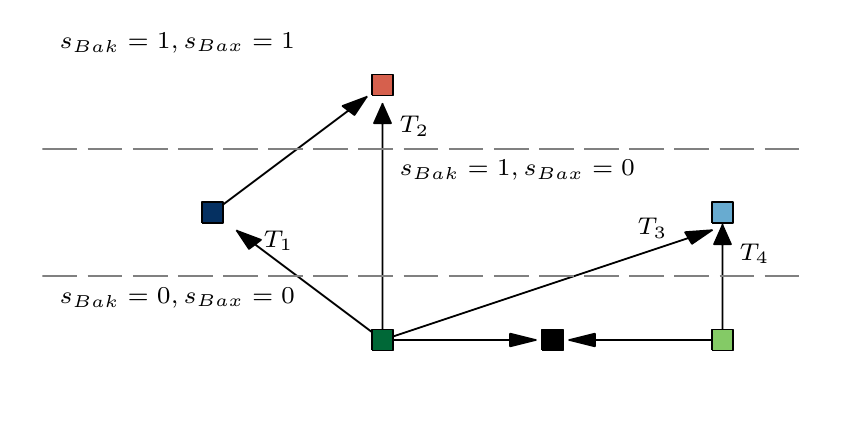}
	\end{center}
	\caption{Possible transitions between the individual subsets of model's ending states.
		 Using the same color notation as in the Fig~\ref{fig:states}, 
		 the types of the model's transitions between five functionally distinct subgroups 
		 of model's states are depicted.}
	\label{fig:transitions}
\end{figure}

When taking previous results together, we may distinguish six distinct
groups of model's ending states. 
Where three of these groups associate survival states, involving
no effectors activity, two of them associate pro-momp states, 
involving the activity of Bak, but not Bax, 
and the one group associating the states involving
the activity of both effectors. 
Assuming that, the occurrence of momp requires 
only the single effector activation, 
four types of survival-to-momp 
transitions can be distinguished 
($T_1$ -- $T_4$, see Fig~\ref{fig:transitions}).

By calculation of the coefficient of multiple correlation 
(for more details see~\ref{appendix}) -- $R^2$, we have analyzed the 
influence of the expression of the given protein on 
the initiation of these transitions. 
In addition to this, we also have studied the importance of the individual proteins' expression,
to obtain the picture about how these affect the momp initiation.

\begin{figure}
	\begin{center}
		\includegraphics[width = 0.4\textwidth]{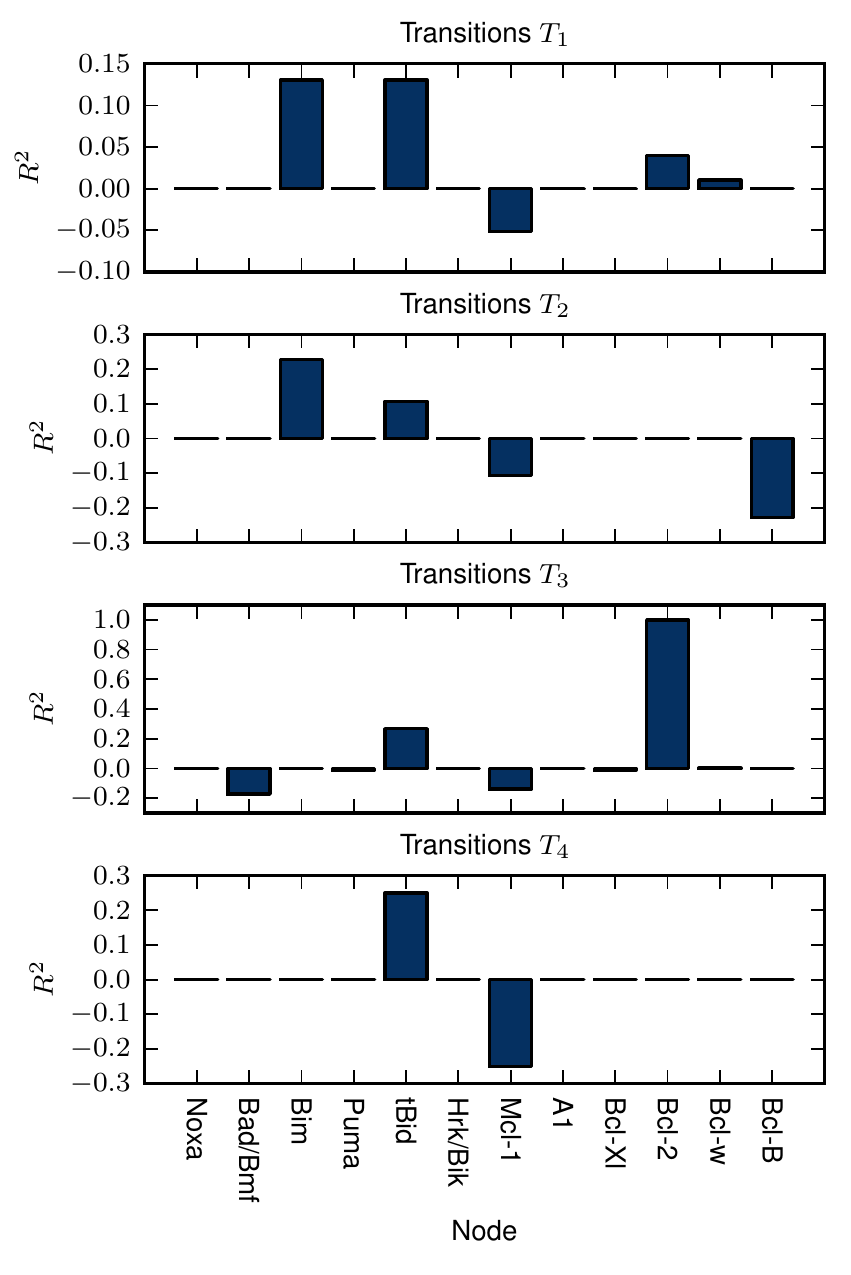}
	\end{center}
	\caption{Multiple determination coefficients -- 
		 $R^2$ of the protein expressions calculated
		 across the sets of unique expressions vectors causing 
		 the survival-to-momp transitions of given type 
		 (for more details see~\ref{appendix}).}
	\label{fig:mcorrels}
\end{figure}

The results we have gained (see Fig~\ref{fig:mcorrels}) show that regardless of the transition type, 
the continuous expression of at least one of the activators (Bim, tBid) is,
as expected,  the key factor determining whether the transition 
to momp occurs or not.
On the other hand it seems that the continuous expression of the 
Mcl-1 is the factor that prevents the momp at the most, compared to
expressions of other antiapoptotic proteins.
It seems that the statistical importance of 
the expression of activators and the absence of Mcl-1 
is common for both $T_1$ and $T_2$.  
Nevertheless, the transitions of type $T_2$ - 
the activation of both effectors, 
additionally relies on the lack of Bcl-B expression.
This finding is not surprisable as the Bcl-B is the 
model's major inhibitor of Bax that is not suppressed by any of the
BH3-only proteins.
Furthermore, model predicts that after the transition 
of the type $T_1$ drives the model to the Bak-active 
pro-momp state, subsequent downregulation of Bcl-B could
cause the additional activation of Bax (the arrow pointing
from the dark blue square to the red one, Fig~\ref{fig:transitions})

Transitions of the type $T_3$ differs from previous types, 
since it occurs in the presence of the Bcl-2 expression, exclusively.
The transitions of type $T_4$ are caused by 
lack of the Mcl-1 expression while involve the tBid expression.
However, the $T_4$ type of transitions may probably occur only rarely,
since the number of "semioptimal" states is very small compared to the number
of "optimal" ones.

Besides the survival-to-momp transitions we have been interested about the
causes of the transition from "optimal"/"semioptimal" to "tumor" states.
We have found (data not shown) that the trapping of the model within the
"tumor" state simply occurs as the anti-apoptotic expression 
disproportionately dominates over the expression of BH3-only proteins.
This points to the necessity of the balance between the presence and synthesis
of the both pro- and anti-apoptotic Bcl-2 proteins within the cell.

\section{Discussion}
\label{sec:discussion}
We have analyzed the Bcl-2 family interaction network using the 
boolean network-based computational model. 
Bcl-2 family members have been represented by model's nodes, which
activity was binary encoded. The active (ON state) of the 
given node represents the biologically active form of the represented
protein. 
Nodes are mutually interacting according to the given transition rules, 
and pre-defined relationship matrix, 
representing the chemical interactions among the Bcl-2 proteins.
In addition, the model behavior has been influenced by the 
vector of expressions, that is representing the biological conditions, 
change of which is the primary driving force of the Bcl-2 regulatory mechanism.
The expression here is a mean to include the post-translational and/or 
post-transcriptional activation of zymogens, if relevant for given protein.

Computational simulations of the model, suggest that 
the Bcl-2 family can withstand wide variety of 
signals controlling the activity of its pro- and anti-apoptotic members,
without causing the activation of Bax or Bak.
While allowing the cell survival, the Bcl-2 family may preserve in 
one of the many different states. 
However, our results show, that once the anti-apoptotic
proteins significantly overwhelm the BH3-only proteins's activity, the 
Bcl-2 family regulation may be seriously harmed. 
Our model predicts, that once this happens, 
even the subsequent activation of pro-apoptotic BH3-only proteins, 
cannot reverse the MOMP regulation malfunctioning.
Moreover, our results suggest that the presence of the anti-apoptotic protein 
Mcl-1 may play an important role here.
However, the existence of such "tumorgeneric" trap shows 
the importance of the equilibrium, dynamically preserved
by the continuous expression of both anti- and pro-apoptotic
proteins.

According to our simulations, depending on the current state of the 
Bcl-2 family, certain configurations of the incoming signals 
can cause state transition including the activation of effectors
Bak, and/or Bax.
As the most important factors among such signals are the 
activation of Bid (its truncation to tBid) and/or the
activation of Bim and the downregulation of Mcl-1.
It seems that the activation of Bak is much less 
stimuli-demanding compared to the activation of Bax. 
However, it seems that the activation of Bax strongly benefits from
the Bcl-w downregulation. 
Finally, our results confirm 
the irreversibility of the effectors activation.

Despite the limited predictive and explanatory power of
the boolean-based approach, we believe that the proposed model
sheds a light on the \textit{modus operandi} of the 
Bcl-2 mediated regulation of momp.
In future we would like to focus our attention on the 
system properties of the same model, including its
robustness and responsivity.

\clearpage
\appendix
\section{Calculation of multiple determination coefficients}
\label{appendix}
This analysis is proposed to 
compare the importance of the presence/absence of particular proteins expression, 
regarding the transition from certain group of states to 
another group of states.
Here we describe the methodology we used in these analyses.

Let's have set of $n$ unique expression vectors -- $E = \{e_1, e_2,\dots e_{12}\}$ 
that cause the studied transition.
For each couple of nodes $i$, $j$ we can calculate the phi coefficient:
\begin{equation}
\phi_{ij} = \frac{n_{11}n_{00} - n_{10}n_{01}}{\sqrt{n_{1\bullet} n_{0\bullet} n_{\bullet0} n_{\bullet1}}},
\end{equation}
where $n_{00}$, $n_{01}$, $n_{10}$, $n_{11}$, are counts of the following combinations
of values $e_i$, $e_j$ across the set of expression vectors:
\begin{table}[ht]
	\footnotesize
	\centering
	\begin{tabular}{cccc}
		& $e_i = 1$ & $e_i = 0$ & total \\
		$e_j = 1$ & $n_{11}$ & $n_{10}$ & $n_{1\bullet}$ \\
		$e_j = 0$ & $n_{01}$ & $n_{00}$ & $n_{0\bullet}$ \\
		total & $n_{\bullet1}$ & $n_{\bullet0}$ & $n$
	\end{tabular}
\end{table}

The phi coefficient is a measure of association for 
two binary variables, similar to Pearson correlation coefficient.

The matrix of phi coefficients - $R_\phi$, is then used as to calculate the 
coefficient of multiple determination -- $R^2$:
\begin{equation}
R^2_i = c^{T}_i R_{\phi, i}^{-1}\,c_i,
\end{equation}
where the $c_i$ is the vector of values $\phi_{ij},\,j = 1, 2\ldots 12,\,i \not= j$.
$c_i$ is actually the vector of correlations between the independent variables and 
the target variable -- $e_i$. $c^{T}_i$ is the transpose of $c$. 
The $R_{\phi, i}$ is the matrix $R_\phi$, reduced by removing the $i$-th line and $i$-th column.
$R_{\phi, i}$ is actually the matrix of correlations between 
the independent variables and $R^{-1}_{\phi, i}$ is the inverse of the matrix $R_{\phi, i}$.

Finally, the $R^2_i$ was multiplied by $-1$ if 
the count of $e_i = 0$ appearances greater than of the $e_i = 1$, across the set of expressions
-- the expression of the $i$-th node was mostly absent among the expression vectors causing the 
given transition.

In the case that the values of $i-th$ protein expression had no variability --
either $e_i = 0$, or $e_i = 1$ among all the expression vectors, 
the $R^2_i$ was arbitrary set either to $-1$, or $1$, respectively. 
The correlations of any of the other expression with 
the $e_i$ were then excluded from any other calculations.
Such situation occurs in case of expression of Bcl-2 among the transitions of the type 
$T_3$ (see Fig~\ref{fig:mcorrels})

\clearpage
\bibliography{Boolean_network-based_model_of_the_Bcl-2_family_mediated_MOMP_regulation}

\begin{thebibliography}{10}
\expandafter\ifx\csname url\endcsname\relax
  \def\url#1{\texttt{#1}}\fi
\expandafter\ifx\csname urlprefix\endcsname\relax\def\urlprefix{URL }\fi
\expandafter\ifx\csname href\endcsname\relax
  \def\href#1#2{#2} \def\path#1{#1}\fi

\bibitem{ulukava_2011}
E.~Ulukava, C.~Acilan, Y.~Y, Apoptosis: why and how does it occur in biology?,
  Cell Biochemistry and Function 29 (2011) 468--80.

\bibitem{wyllie_2010}
A.~H. Wyllie, "where, o death, is thy sting?" a brief review of apoptosis
  biology, Molecular Neurobiology 42 (2010) 4--9.

\bibitem{elmore_2007}
S.~Elmore, Apoptosis: A review of programmed cell death, Toxicologic Pathology
  35 (2007) 495--516.

\bibitem{mattson_2006}
M.~P. Mattson, Neuronal life-and-death signaling, apoptosis, and
  neurodegenerative disorders, Antioxidants \& Redox Signaling 8.

\bibitem{nagata_autoimmune_2010}
S.~Nagata, Apoptosis and autoimmune diseases, Annals of the New York Academy of
  Sciences 1209 (2010) 10--16.

\bibitem{burz_apoptosis_2009}
C.~Burz, I.~Berindan-Neagoe, O.~Balacescu, A.~Irimie, Apoptosis in cancer: Key
  molecular signaling pathways and therapy targets, Acta Oncologica 48 (2009)
  811--821.

\bibitem{fulda_2009}
S.~Fulda, Tumor resistance to apoptosis, International Journal of Cancer 124
  (2009) 511--515.

\bibitem{strasser_deciphering_2011}
A.~Strasser, S.~Cory, J.~M. Adams, Deciphering the rules of programmed cell
  death to improve therapy of cancer and other diseases, EMBO Journal 30 (2011)
  3667--3683.

\bibitem{strasser_apoptosis_2000}
A.~Strasser, L.~O'Connor, V.~M. Dixit, Apoptosis signaling, Annual Review of
  Biochemistry 69 (2000) 217--245.

\bibitem{danial_2004}
N.~N. Danial, S.~J. Korsmeyer, Cell death: Critical control points, Cell 116
  (2004) 205--219.

\bibitem{chipuk_2008}
J.~E. Chipuk, D.~R. Green, How do bcl-2 proteins induce mitochondrial outer
  membrane permeabilization?, Trends in Cell Biology 18 (2008) 157--164.

\bibitem{tait_2010}
S.~W.~G. Tait, D.~R. Green, Mitochondria and cell death: outer membrane
  permeabilization and beyond, Nature Reviews Molecular Cell Biology 11 (2010)
  621--632.

\bibitem{landes_2011}
T.~Landes, J.~C. Martinou, Mitochondrial outer membrane permeabilization during
  apoptosis: the role of mitochondrial fission, Biochimica et Biophysica Acta
  1813 (2011) 540--545.

\bibitem{mace_2010}
P.~D. Mace, S.~J. Riedl, Molecular cell death platforms and assemblies, Current
  Opinion In Cell Biology 22 (2010) 828--836.

\bibitem{perez-paya_2011}
E.~Perez-Paya, M.~Orzaez, L.~Mondragon, D.~Wolan, J.~A. Wells, A.~Messequer,
  M.~J. Vincent, Molecules that modulate apaf-1 activity, Medical Research
  Reviews 31 (2011) 649--675.

\bibitem{kulikov_2012}
A.~V. Kulikov, E.~S. Shilov, I.~A. Mufazalov, V.~Gogvadze, S.~A. Nedospasov,
  B.~Zhivotinsky, Cytochrome c: the achilles' heel in apoptosis, Cellular and
  Molecular Life Sciences 69 (2012) 1787--1797.

\bibitem{wurstle_2012}
M.~L. Wurstle, M.~A. Laussmann, M.~Rehm, The central role of initiator
  caspase-9 in apoptosis signal transduction and the regulation of its
  activation and activity on the apoptosome, Experimental Cell Research 318
  (2012) 1213--1220.

\bibitem{martinez-ruiz_2008}
G.~Martinez-Ruiz, V.~Maldonado, G.~Caballos-Cancino, J.~P. Grajeda,
  J.~Melendez-Zajgla, Role of smac/diablo in cancer progression, Journal of
  Experimental \& Clinical Cancer Research 27.

\bibitem{olsson_2011}
M.~Olsson, B.~Zhivotinsky, Capases and cancer, Cell Death \& Differentiation 18
  (2011) 1441--1449.

\bibitem{green_1998}
D.~R. Green, G.~B. Amarate-Mendes, The point of no return: mitochondria,
  caspases, and the commitment to cell death, Results and Problems in Cell
  Differentiation 24 (1999) 45--61.

\bibitem{wallach_2008}
D.~Wallach, T.-B. Kang, A.~Kovalenko, The extrinsic cell death pathway and the
  elan mortel, Cell Death \& Differentiation 15 (2008) 1533--1541.

\bibitem{estaquier_2012}
J.~Estaquier, F.~Vallette, J.~L. Vayssiere, B.~Mignotte, The mitochondrial
  pathways of apoptosis, Advances in Experimental Medicine and Biology 942
  (2012) 157--183.

\bibitem{chen_robustness_2007}
C.~Chen, J.~Cui, W.~Zhang, P.~Shen, Robustness analysis identifies the
  plausible model of the bcl-2 apoptotic switch, FEBS Letters 581 (2007)
  5143--5150.

\bibitem{cui_two_2008}
J.~Cui, C.~Chen, H.~Lu, T.~Sun, P.~Shen, Two independent positive feedbacks and
  bistability in the bcl-2 apoptotic switch, PLoS ONE 3 (2008) 1469.

\bibitem{sun_2010}
T.~Sun, X.~Lin, Y.~Wei, Y.~Xu, P.~Shen, Evaluating bistability of bax
  activation switch, FEBS Letters 584 (2010) 954--960.

\bibitem{tokar_2012}
T.~Tokar, J.~Ulicny, Computational study of bcl-2 apoptotic switch, Physica A.

\bibitem{bagci_2006}
E.~Z. Bagci, Y.~Vodovotz, T.~R. Billiar, G.~B. Ermentrout, I.~Bahar,
  Bistability in apoptosis: Roles of bax, bcl-2, and mitochondrial permeability
  transition pores, Biophysical Journal 90 (2006) 1546--1559.

\bibitem{albeck_quantitative_2008}
J.~G. Abeck, J.~M. Burke, B.~B. Aldridge, M.~Zhang, D.~A. Lauffenburger, P.~K.
  Sorger, Quantitative analysis of pathways controlling extrinsic apoptosis in
  single cells, Molecular Cell 30 (2008) 11--25.

\bibitem{harrington_2008}
H.~Harrington, K.~L. Lo, S.~Ghosh, K.~Tung, Construction and analysis of a
  modular model of caspase activation in apoptosis, Theoretical Biology and
  Medical Modelling 90 (2008) 1546--1559.

\bibitem{schlatter_2009}
R.~Schlatter, K.~Schmich, I.~Avalos~Vizcarra, P.~Scheurich, T.~Sauter,
  C.~Borner, M.~Ederer, I.~Merfort, O.~Sawodny, On/off and beyond--a boolean
  model of apoptosis, PLoS Computational Biology 5.

\bibitem{mai_boolean_2009}
Z.~Mai, H.~Liu, Boolean network-based analysis of the apoptosis network:
  Irreversible apoptosis and stable surviving, Journal of Theoretical Biology
  259 (2009) 760--769.

\bibitem{kauffman_1969}
S.~A. Kauffman, Metabolic stability and epigenesis in randomly constructed
  genetic nets, Journal of Theoretical Biology 22 (1969) 437--467.

\bibitem{helikar_2011}
T.~Helikar, N.~Kochi, J.~Konvalina, J.~A. Rogers, Boolean modeling of
  biochemical networks, The Open Bioinformatics Journal 5 (2011) 16--25.

\bibitem{calzolari_2007}
D.~Calzolari, D.~Paternostro, P.~L.~J. Harrington, C.~Piermarocchi, P.~M.
  Duxbury, Selective control of the apoptosis signaling network in heterogenous
  cell populations, PLoS One 2.

\bibitem{elkholi_2011}
R.~Elkholi, K.~V. Floros, J.~E. Chipuk, The role of bh3-only proteins in tumor
  cell development, signaling, and treatment, Genes Cancer 2.

\bibitem{chipuk_reunion_2010}
J.~E. Chipuk, T.~Moldoveanu, F.~Llambi, M.~J. Parsons, D.~R. Green, The bcl-2
  family reunion, Molecular Cell 37 (2010) 299--310.

\bibitem{placzek_2010}
W.~J. Placzek, J.~Wei, S.~Kitada, D.~Zhai, J.~C. Reed, M.~Pellecchia, A survey
  of the anti-apoptotic bcl-2 subfamily expression in cancer types provides a
  platform to predict the efficacy of bcl-2 antagonists in cancer therapy, Cell
  Death and Disease 6.

\bibitem{westphal_2011}
D.~Westhpal, G.~Dewson, P.~E. Czabotar, R.~M. Kluck, Molecular biology of bax
  and bak activation and action, Biochimica Biophysica Acta 4 (2011) 521--531.

\bibitem{dejean_2010}
L.~M. Dejean, S.~Y. Ryu, S.~Martinez-Caballero, O.~Teijido, P.~M. Peixoto,
  K.~W. Kinnally, Mac and bcl-2 family proteins conspire in a deadly plot,
  Biochimica Biophysica Acta 1797 (2010) 1231--1238.

\bibitem{chen_differential_2005}
L.~Chen, S.~N. Willis, A.~Wei, B.~J. Smith, J.~I. Fletcher, M.~G. Hinds, P.~M.
  Colman, C.~L. Day, J.~M. Adams, D.~C.~S. Huang, Differential targeting of
  prosurvival bcl-2 proteins by their bh3-only ligands allows complementary
  apoptotic function, Molecular Cell 17 (2005) 393--403.

\bibitem{rautureau_2010}
G.~J.~P. Rautureau, C.~L. Day, M.~G. Hinds, The structure of boo/diva reveals a
  divergent bcl-2 protein, Proteins 78 (2010) 2181--2186.

\bibitem{kantari_2011}
C.~Kantari, H.~Walczak, Caspase-8 and bid: caught in the act between death
  receptors and mitochondria, Biochimica Biophysica Acta 4 (2011) 558--563.

\end{thebibliography}
\bibliographystyle{elsarticle-num}

\end{document}